\documentclass[fleqn,10pt]{wlscirep}
\usepackage[utf8]{inputenc}
\usepackage[T1]{fontenc}
\usepackage{bm}
\usepackage{comment}

\title{Current-Controlled Nanomagnetic Writing for Reconfigurable Magnonic Crystals}

\author[1,*]{J. C. Gartside}
\author[1,2]{S. G. Jung}
\author[1]{S. Y. Yoo}
\author[3]{D. M. Arroo}
\author[1]{A. Vanstone}
\author[1,3]{T. Dion}
\author[1]{K. D. Stenning}
\author[1]{W. R. Branford}
\affil[1]{Blackett Laboratory, Imperial College London, London SW7 2AZ, United Kingdom}
\affil[2]{Now at Cavendish Laboratory, University of Cambridge, Cambridge, CB3 0HE, United Kingdom}
\affil[3]{London Centre for Nanotechnology, University College London, London WC1H 0AH, United Kingdom}
\affil[*]{Corresponding author e-mail: j.carter-gartside13@imperial.ac.uk}

\begin{abstract}

Strongly-interacting nanomagnetic arrays are crucial across an ever-growing suite of technologies. Spanning neuromorphic computing, control over superconducting vortices and reconfigurable magnonics, the utility and appeal of these arrays lies in their vast range of distinct, stable magnetisation states. Different states exhibit different functional behaviours, making precise, reconfigurable state control an essential cornerstone of such systems. However, few existing methodologies may reverse an arbitrary array element, and even fewer may do so under electrical control, vital for device integration.

We demonstrate selective, reconfigurable magnetic reversal of ferromagnetic nanoislands via current-driven motion of a transverse domain wall in an adjacent nanowire. The reversal technique operates under all-electrical control with no reliance on external magnetic fields, rendering it highly suitable for device integration across a host of magnonic, spintronic and neuromorphic logic architectures. Here, the reversal technique is leveraged to realise two fully solid-state reconfigurable magnonic crystals, offering magnonic gating, filtering, transistor-like switching and peak-shifting without reliance on global magnetic fields.
\end{abstract}

\begin{document}

\flushbottom
\maketitle
\thispagestyle{empty}

\section*{Introduction}

Versatile, low-power means to selectively control magnetisation states at the nanoscale are critical across a host of applications, both in fundamental science and device-oriented systems. Alongside mature technologies such as data storage, nanomagnetic arrays support a host of more recent applications including neuromorphic computation\cite{romera2018vowel,mizrahi2018neural,grollier2020neuromorphic,sangwan2020neuromorphic}, superconducting vortex control\cite{wang2018switchable,sadovskyy2017effect,rollano2019topologically} and reconfigurable magnonic crystals\cite{grundler2015reconfigurable,haldar2016reconfigurable,krawczyk2014review,wang2017voltage,topp2010making} (RMCs). RMCs are nanopatterned metamaterials harnessing varying magnetic configurations to manipulate and store information by tuning magnonic (spin-wave) dynamics\cite{neusser2009magnonics,kruglyak2010magnonics,khitun2010magnonic,khitun2011non,lenk2011building,chumak2015magnon,chumak2017magnonic,wang2019integrated,fischer2017experimental,wintz2016magnetic,sluka2019emission,liu2019current}.

RMCs promise to be key functional elements in the burgeoning field of magnonics\cite{kruglyak2010magnonics,lenk2011building,khitun2010magnonic,khitun2011non,chumak2017magnonic,wang2017voltage,wang2019integrated}. Comprising arrays of interacting nanomagnets, different magnetisation configurations of the array (or `microstates') allow a range of functional behaviours including amplifying and attenuating specific frequency channels\cite{sadovnikov2018spin,wang2015tunable,semenova2013spin,ma2011micromagnetic,kim2009gigahertz} or bandgap creation and tuning\cite{mamica2019reversible,mamica2019nonuniform,graczyk2018magnonic}.
Existing magnonic crystal designs have shown strong initial promise, but are severely limited by a lack of practical means to access more than a handful of microstates. N-element arrays possess $2^N$ microstates, and typically systems with $2^{10} - 2^{100}$ potential microstates are restricted to operating within just two or three configurations\cite{haldar2016reconfigurable}, imposing a hard limit on scope and utility.

Of the existing microstate control techniques, from simple applications of system-wide magnetic fields to Oersted-field stripline techniques and intricate multilayered spin-transfer torque devices\cite{kawahara20072mb,zhao2009spin,kawahara20082,kawahara2012spin}, few are able to selectively reverse the magnetisation of an arbitrary element in a strongly-interacting nanoarray without affecting neighbouring elements. Many approaches struggle to function without disturbing delicate magnetic states elsewhere in the system due to large stray fields or a reliance on global external fields. The remaining methods rely on mechanical apparatus orders of magnitude larger than the nanomagnet, either a hard-disk style write-head or scanning-probe with magnetic tip\cite{gartside2018realization,wang2016rewritable,gartside2016novel} --- unsuitable for on-chip integration and susceptible to damage.

\begin{figure}[htbp]
\centering
\includegraphics[width=0.99\textwidth]{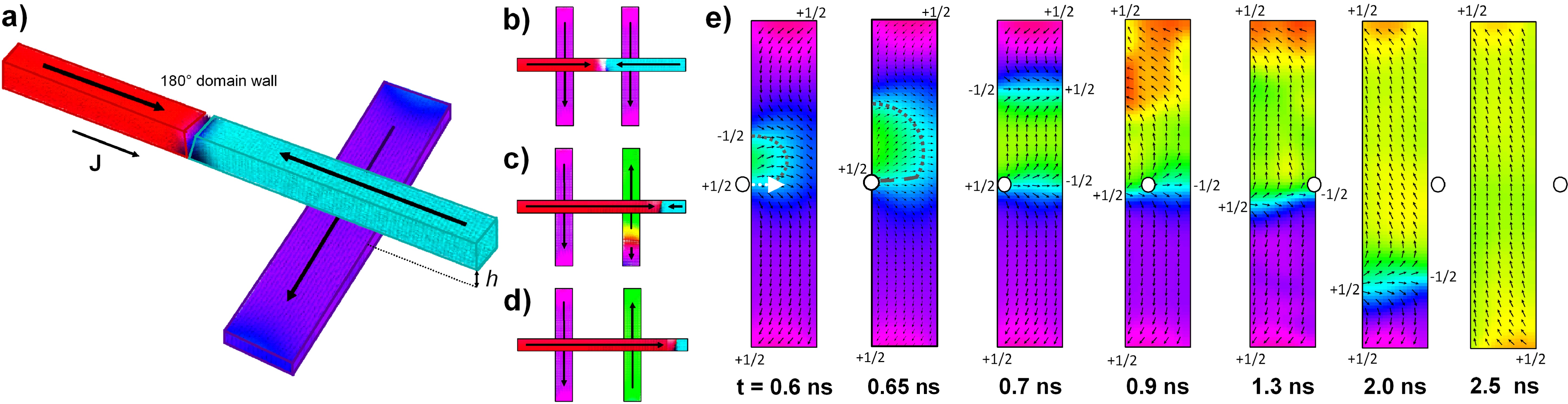}
\caption{Schematic of the magnetic reversal method. \\ 
\textbf{a)} Representation of the ferromagnetic DW-carrying `control' nanowire and `bit' nanoisland. \\
\textbf{b-d)} Time-evolution series of the magnetic reversal process in two-nanoisland system. Both nanoislands are initially magnetised with $M = -\hat{y}$ (\textbf{b}) before the current-driven `control' domain wall (c-DW) traverses the right-hand nanoisland, inducing magnetisation reversal (\textbf{c}), leaving it magnetised $M = +\hat{y}$ while the left-hand nanoisland remains unswitched (\textbf{d}). \\
\textbf{e)} Detailed time-evolution series of the reversal process in a single $400 \times 75 \times 5~$ nm$^3$ nanoisland. The dynamic c-DW is represented by the white circle traversing the nanoisland midpoint, moving with $v_{\mathrm{c-DW}} = +\hat{x}$. $t = 0$~s is defined when the c-DW begins moving from its origin, 112 nm to the nanoisland's left. The partially-formed contorted nanoisland DW is highlighted in $t = 0.6$~ns by the dashed grey line, with the partially-formed straight island DW highlighted by longer grey dashes in $t = 0.65$~ns. Topological defects are labelled with their winding numbers. \\
Corresponding time-evolution series are provided in supplementary videos SV1 and SV2.}
\label{Fig1} 
\end{figure}

To address the pressing need for non-invasive, global-field free means for nanomagnetic writing, we build on the fundamental principles of `topological magnetic writing'\cite{gartside2018realization} to present a scheme enabling low-power, fully solid-state access to the entire microstate space of strongly-interacting nanomagnetic systems. 
The tightly-localised stray-field of a current-driven 180$^{\circ}$ transverse domain wall (DW) in a nanowire is used to induce dynamic topological defects in adjacent ferromagnetic nanoislands, driving magnetic reversal. By varying the drive-current amplitude in the nanowire, fully selective reversal is achieved and nanoislands may be switched or `skipped' at will as the DW passes.
The versatility and utility of the technique are demonstrated via two novel active magnonic systems; a reconfigurably-gateable 1D transmission-line RMC optimised for travelling-wave magnons supporting multiple gate types and on/off ratios up to 35, and a 2D RMC optimised for standing-wave magnons with single-frequency on/off ratios of up to $ 9 \times 10^3$ and mode shifting of $\Delta f = 0.96~$ GHz. 

\section*{Working principle of reversal method} 

The reversal process is depicted schematically in Fig. \ref{Fig1}a-d) with a corresponding time evolution series shown in Fig. \ref{Fig1} e). The system comprises a ferromagnetic DW-carrying `control' nanowire (here Permalloy (Py)) and Ising-like Py `bit' nanoisland(s) at a height $h$ below the control wire. The control wire serves as a track allowing a transverse control-DW (hereafter c-DW) to traverse the nanoislands via current-induced spin-transfer torque. The c-DW stray field influences the nanoisland magnetisations as it moves over them, allowing the islands to serve as rewritable bits. N.B. the c-DW nomenclature serves just to distinguish the control nanowire DW from DWs in the bit nanoislands. All DWs discussed are standard transverse DWs. \\

Figure \ref{Fig1} e) shows a detailed micromagnetic time-evolution of the reversal process. Here we view a single Py nanoisland from above (positive $z$-direction, nanoisland in the $xy$-plane) as traversed by a c-DW (white circle) in the Py control nanowire, moving with $v_{\mathrm{c-DW}} = +\hat{x}$. The nanoisland is initially magnetised with $M = -\hat{y}$ and the chirality of the c-DW such that its magnetisation at the c-DW centre points in the $-\hat{z}$ direction towards the nanoisland. The nanoisland dimensions are $400 \times 75 \times 5~$ nm$^3$ and the nanowire has quasi-infinite length, $40 \times 40~$ nm$^2$ cross-section and suspension height $h = 10~$ nm above the nanoisland. 

As the c-DW approaches the nanoisland (fig. \ref{Fig1}~e), $t$ = 0.6 ns) the c-DW stray-field $H_{\mathrm{DW}}$ distorts the spins on the left-edge of the nanoisland, forcing them out of a collinear state to lie along the locally-divergent radial $H_{\mathrm{DW}}$. This distortion of the nanoisland magnetisation introduces a pair of edge-bound topological defects with opposite polarity $\pm 1/2$ winding numbers\cite{tchernyshyov2005fractional}, seen on the nanoisland left edge. Winding numbers in a ferromagnet are conserved and must sum to 1 in a hole-free nanoisland. As the c-DW progresses across the nanoisland ($t$ = 0.65 ns), the influence of $H_{\mathrm{DW}}$ extends to the nanoisland's right-edge, eventually causing a corresponding pair of edge-bound $\mp 1/2$ topological defects to form ($t$ = 0.7 ns). 

Continuous chains of reversed spins now connect each topological defect pair across the nanoisland width, effectively binding the $\pm 1/2$ defects together via the exchange-energy penalty for breaking the spin-chains. Each bound defect-pair constitutes a 180$^{\circ}$ DW, however the two nanoisland DWs (hereafter i-DWs) are formed asymmetrically. 
The lower ($y$-direction) i-DW directly under the c-DW path (dashed white arrow in panel $t$ = 0.6 ns) forms in its lowest-energy straight conformation parallel to the nanoisland width, as $H_{\mathrm{DW}}$ is locally oriented along this axis. For the upper i-DW, $H_{\mathrm{DW}}$ is oriented in the positive $y$-direction antiparallel to the initial nanoisland magnetisation. This forces the nascent upper i-DW to assume a contorted conformation around the growing $M = +\hat{y}$ domain (yellow-green region in fig. \ref{Fig1}e)), with a corresponding exchange-energy penalty relative to the straight lower i-DW. The difference in nascent i-DW conformations is seen in $t$ = 0.65 ns, with the lower, straight and upper, contorted i-DWs highlighted by long and short grey dashed lines respectively.

Once the c-DW progresses far enough to introduce topological edge-defects to the nanoisland's far-side, the $M = +\hat{y}$ domain is fully-formed and the contorted i-DW no longer forced to assume a high-energy conformation. As such it rapidly straightens out, assuming a low-energy conformation straight across the nanoisland width and converting its excess exchange energy into motion in the process. The released exchange energy propels the i-DW along the nanoisland in the positive $y$-direction until it collides with the nanoisland's upper end. Upon reaching the island end the $\pm 1/2$ defects are free to reach and annihilate each other, unwinding the i-DW to a collinear state and emitting a spin-wave burst down the nanoisland ($t$ = 0.9 ns). The spin-wave burst interacts with the remaining i-DW\cite{wang2012domain,gartside2018realization,gartside2016novel}, exerting a torque that accelerates it towards the nanoisland bottom ($t$ = 1.3 ns). As the remaining i-DW approaches the nanoisland bottom it is increasingly magnetostatically attracted to it, further driving acceleration ($t$ = 2.0 ns) until the i-DW unwinds on contact as described above, leaving a DW-free nanoisland magnetised antiparallel to its initial state ($t$ = 2.5 ns) with reversal completed $\sim$ 2 ns after the c-DW reaches the nanoisland.

\begin{figure*}[htbp]   
\centering
\includegraphics[width=0.9\textwidth]{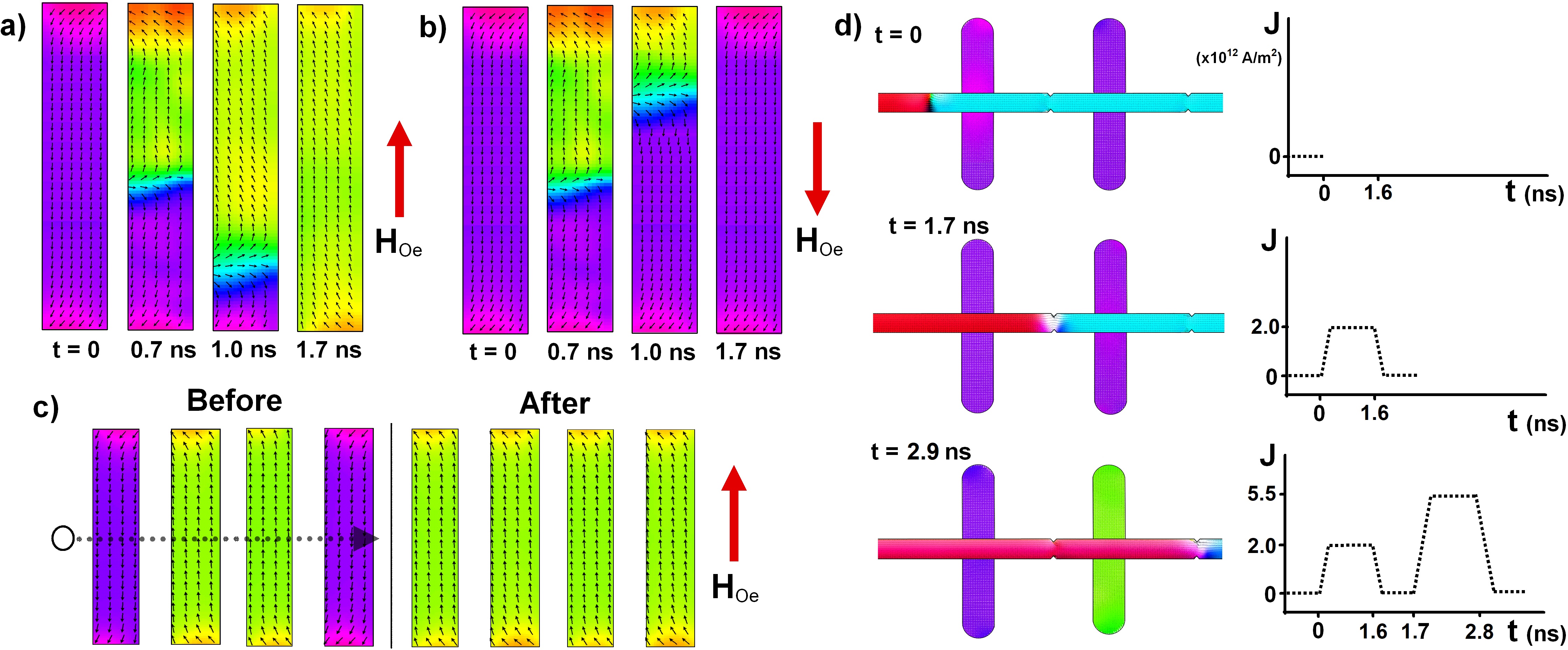}
\caption{Time-evolution series (\textbf{a,b}) and array schematic (\textbf{c}) of close-proximity reversal mode and time-evolution series (\textbf{d}) of fully-selective reversal including current-density $J$ vs. time plots. \\
\textbf{a)} With Oersted-field $H_{\mathrm{Oe}}$ antiparallel to the initial nanoisland magnetisation $H_{\mathrm{Oe}}$ aids reversal, driving the low- and high-energy nanoisland DWs (i-DWs) to opposite nanoisland ends and achieving successful switching. \\
\textbf{b)} With $H_{\mathrm{Oe}}$ aligned to the initial nanoisland magnetisation the reversal process is hindered, $H_{\mathrm{Oe}}$ drives the low- and high-energy i-DWs into the same nanoisland end, resulting in failed switching. \\
\textbf{c)} Four-island array initialised (`before' panel) with second and third islands in $M = +\hat{y}$ state and first and fourth in $M = -\hat{y}$ state. Control-DW (c-DW) is then driven across the array such that $H_{\mathrm{Oe}} = +\hat{y}$, driving successful reversal of $M = -\hat{y}$ islands (as described in panel a) and failed reversal of $M = +\hat{y}$ islands (as in panel b).  
\textbf{d)} c-DW begins to the left of two nanoislands (t = 0). It is driven over the first nanoisland by a $J = 2.0 \times 10^{12}$ A/m$^2$ `skip' current pulse, leaving the nanoisland unswitched as $H_{\mathrm{Oe}} + H_{\mathrm{DW}} < H_{\mathrm{N}}$, the nucleation field. 
The c-DW is then driven over the second nanoisland with a $J = 5.5 \times 10^{12}$ A/m$^2$ `write' pulse, switching the nanoisland magnetisation as $H_{\mathrm{Oe}} + H_{\mathrm{DW}} \geq H_{\mathrm{N}}$.
} 
\label{Fig3} \vspace{-1em}
\end{figure*}

\section*{Distinct modes of reversal method}

The process above describes the core physics behind the reversal process, however the Oersted-field $H_{\mathrm{Oe}}$ arising from the drive-current through the control nanowire has not yet been considered. In the absence of $H_{\mathrm{Oe}}$ (for instance if the c-DW was driven via global $B$ field), the reversal process occurs as described previously for all permutations of Ising-like initial nanoisland magnetisations ($M =\pm \hat y$) and c-DW trajectories ($v_{\mathrm{c-DW}} =\pm \hat x$). However, reversing the c-DW velocity when driving via current requires reversing the current direction and therefore the direction of $H_{\mathrm{Oe}}$ in the nanoisland plane ($\pm y$ direction). 
This leads to two distinct reversal regimes occurring at different nanowire suspension heights; a fully-selective mode where $h = 20-25~$ nm, allowing reversal of any desired element in a nanoarray, and a close-proximity mode allowing nanoisland reversal at lower current-density and higher bit-density than $H_{\mathrm{Oe}}$ stripline-based techniques occurs where $h = 2-11~$ nm, albeit at reduced selectivity relative to the $h = 20-25~$ nm mode. Nanoislands of dimensions $400 \times 75 \times 5~$ nm$^3$ are considered in both reversal regimes. 

\subsection*{Close-proximity, low-power reversal}

Considering first the close-proximity mode, here the contorted high-energy i-DW is injected with sufficient energy to overcome the influence of both possible $\pm \hat y$ $H_{\mathrm{Oe}}$ directions, and will unwind into its nearest nanoisland end as the c-DW passes as in fig.\ref{Fig1}e) $t$ = 0.7-0.9 ns. However, the behaviour of the initially static low-energy i-DW is determined by the drive-current polarity and corresponding $H_{\mathrm{Oe}}$ direction. Fig. \ref{Fig3} shows the low-energy i-DW driven `favourably' (a) by an $H_{\mathrm{Oe}}$ aligned antiparallel to the initial nanoisland magnetisation --- unwinding at the opposite nanoisland end as the high-energy i-DW and successfully mediating reversal, or `unfavourably' (b) by an $H_{\mathrm{Oe}}$ parallel with the initial island magnetisation --- unwinding at the same end as the high-energy i-DW, leaving an unswitched nanoisland magnetised along its initial direction. 

Successful reversal therefore requires an $H_{\mathrm{Oe}}$ aligned antiparallel to the initial nanoisland magnetisation. This means that in the close-proximity regime microstates may be written from a collinear initial state (all nanoislands identically magnetised) such that every nanoisland traversed by the c-DW is switched, with the halting point of the c-DW decided by the user. For instance in a field-saturated ten-island array with c-DW initialised left of the array, island 1 could be switched, or islands 1-5 - but island 5 could not be switched without reversing islands 1-4. For the 400 $\times$ 75 $\times$ 5 nm$^3$ nanoislands considered here the close-proximity mode functions with a minimum current density of $\mathbf{J} = 3 \times 10^{12}$ A/m$^2$ for 5 nm thick Py nanoislands, a factor of $2.3\times$ lower than the minimum $J$ required for conventional non-selective $H_{\mathrm{Oe}}$ reversal of matching nanoisland and nanowire dimensions. Since power consumption and Joule heating scale as $J^2$, c-DW driven reversal leads to an 82\% reduction in power consumption compared to using only the Oersted field while still affording partial selectivity.

The upper bound of $h =$ 11 nm is given for close-proximity reversal as for lower separations, the minimum $J$ required to overcome magnetostatic attraction between c-DW and nanoisland and move the c-DW across the island will also generate sufficient $H_{\mathrm{Oe}}$ to switch the nanoisland, negating `skip' events where the c-DW passes a nanoisland without switching it and hence preventing full selectivity. Increments of 0.5 nm were considered for $h$ down to a minimum of $h =$ 2 nm.

\subsection*{Fully-selective reversal}

In the fully-selective $h = 20-25$ nm regime, the increased nanowire-nanoisland separation means the magnitude of the c-DW stray field $H_{\mathrm{DW}}$ at the nanoisland is lower than the critical nucleation field $H_{\mathrm{N}}$ required to overcome exchange energy and begin locally rearranging spins to nucleate a nascent domain, hence $H_{\mathrm{DW}}$ alone is unable to drive reversal. However, when combined with a sufficient magnitude $H_{\mathrm{Oe}}$ such that $H_{\mathrm{DW}} + H_{\mathrm{Oe}} \geq H_{\mathrm{N}}$, reversal occurs as described above. This allows for two current-driving operations to be performed as shown in fig. \ref{Fig3} d): a `skip' event (t = 1.7 ns), where a low current-density pulse $\textbf{J} = 1.5-3.0 \times 10^{12}$ A/m$^2$ is used to move the c-DW over a nanoisland while maintaining an $H_{\mathrm{Oe}}$ magnitude below the reversal threshold, and a `write' event (t = 2.9 ns) where a higher current-density pulse $\textbf{J} = 5.0-7.0 \times 10^{12}$ A/m$^2$ is used to move the c-DW while providing a sufficiently high $H_{\mathrm{Oe}}$ such that magnetic reversal occurs (Fig. \ref{Fig3} d), t = 1.4 ns). By combining sequences of these `skip' and `write' events, any desired microstate may be realised. To allow for reliable position-control of the c-DW, 5 nm notches were inserted into the control nanowire between island positions (seen in fig. \ref{Fig3} d)), providing local potential wells to ensure that c-DW traverses only one island per pulse and provide a well-defined end position for the c-DW after each pulse.  \\

The threshold current values were determined using simulation increments of $0.5 \times 10^{12}$ A/m$^2$ and defined as follows; `skip' mode: above minimum $J$ to overcome magnetostatic attraction between c-DW and nanoisland and drive the c-DW across the nanoisland, and below maximum $J$ satisfying $H_{\mathrm{DW}} + H_{\mathrm{Oe}} < H_{\mathrm{N}}$ --- and `write' mode: above minimum $J$ satisfying $H_{\mathrm{DW}} + H_{\mathrm{Oe}} \geq H_{\mathrm{N}}$ and below maximum $J$ where induced nanoisland-DWs leave a non-turbulent regime and begin exhibiting Walker breakdown, which can cause stochastic reversal failure by driving both i-DWs into the same nanoisland end.
Functionality of both reversal modes in nanoisland arrays was investigated, with both modes acheiving switching as described above down to a minimum inter-island spacing of 60 nm, below which dipolar coupling between islands begins to interfere with the reversal process.

Existing solid-state magnetic-reversal schemes require two control lines to address a single element in an array; typically two orthogonal sets of current-lines above and below the array\cite{kawahara20072mb,zhao2009spin,kawahara20082,kawahara2012spin}. Here the same control is achieved with a single set of lines, a reduction of $n$ lines in an $n$-row array. 
The c-DW may be prepared electrically using previously described stripline-based schemes\cite{pushp2013domain,burn2017dynamic}.

\begin{figure*}[htbp]
\centering
\includegraphics[width=1\textwidth]{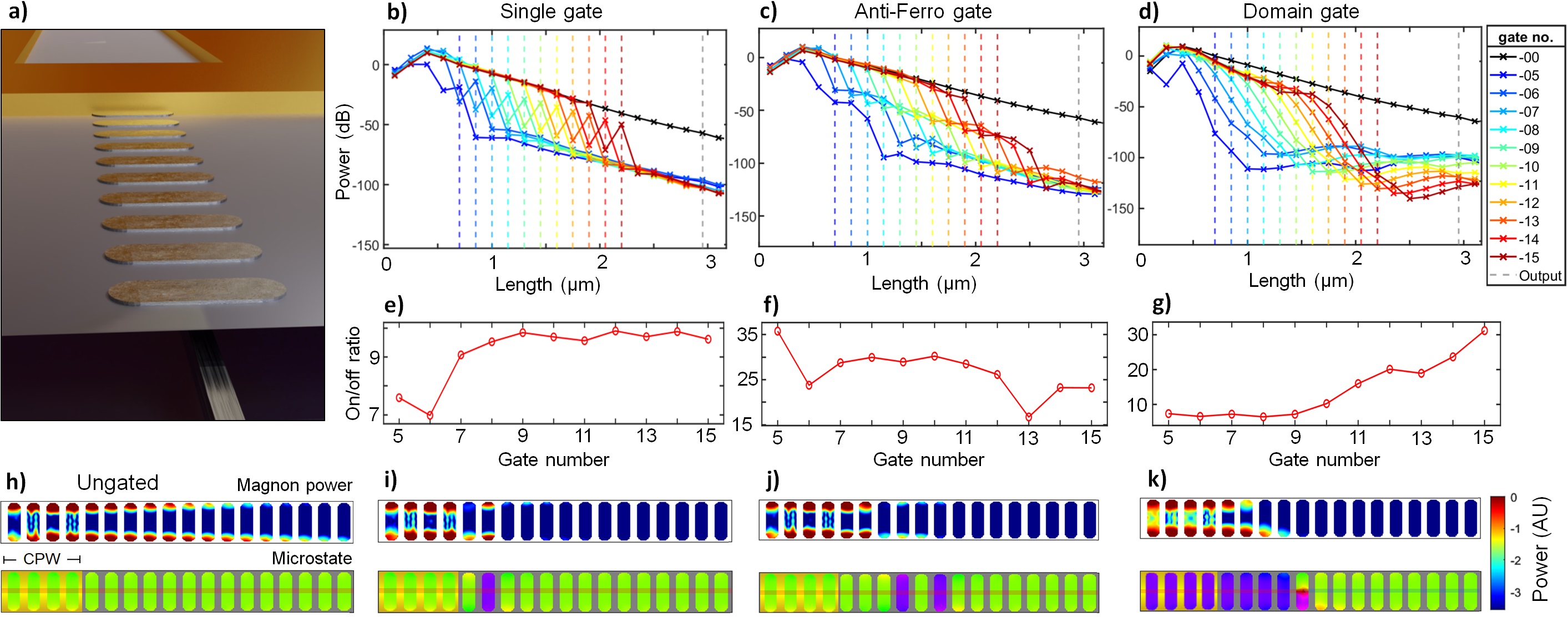}
\caption{a) Schematic of the one-dimensional reconfigurable magnonic crystal. Coplanar Au microwave waveguide is at image top and underlying DW-carrying control nanowire is shown protruding at image bottom. \\
b-d) Edge-mode spin-wave power vs. array position for (b) single, (c) `antiferro' and (d) `domain' gate types. Coplanar microwave waveguide covers islands 1-4. Nanoisland positions denoted by dashed vertical lines with power vs. position traces colour-coded by gate position. \\ 
e-f) Corresponding magnon `on/off' ratio vs. gate position for each gate type, calculated from the ratio of integrated power at island 20 in the ungated to gated case. \\ 
h-k) Spatial power maps of resonant magnon edge mode for (h) ungated, (i) single, (j) `antiferro' and (k) `domain' gate types. Power is normalised to ungated case. Array microstate is shown with green (purple) representing unswitched (switched) islands magnetised in the positive-y (negative-y) direction. Control nanowire is shown over array, coplanar Au waveguide covers islands 1-4. \\
Control-nanowire is in fully-selective $h = 20$~nm  reversal regime for all cases other than `domain gate', which operates in low-power $h = 10$~nm regime.
}
\label{Fig4} \vspace{-1em}
\end{figure*}

\begin{figure*}[htbp]
\centering
\includegraphics[width=1\textwidth]{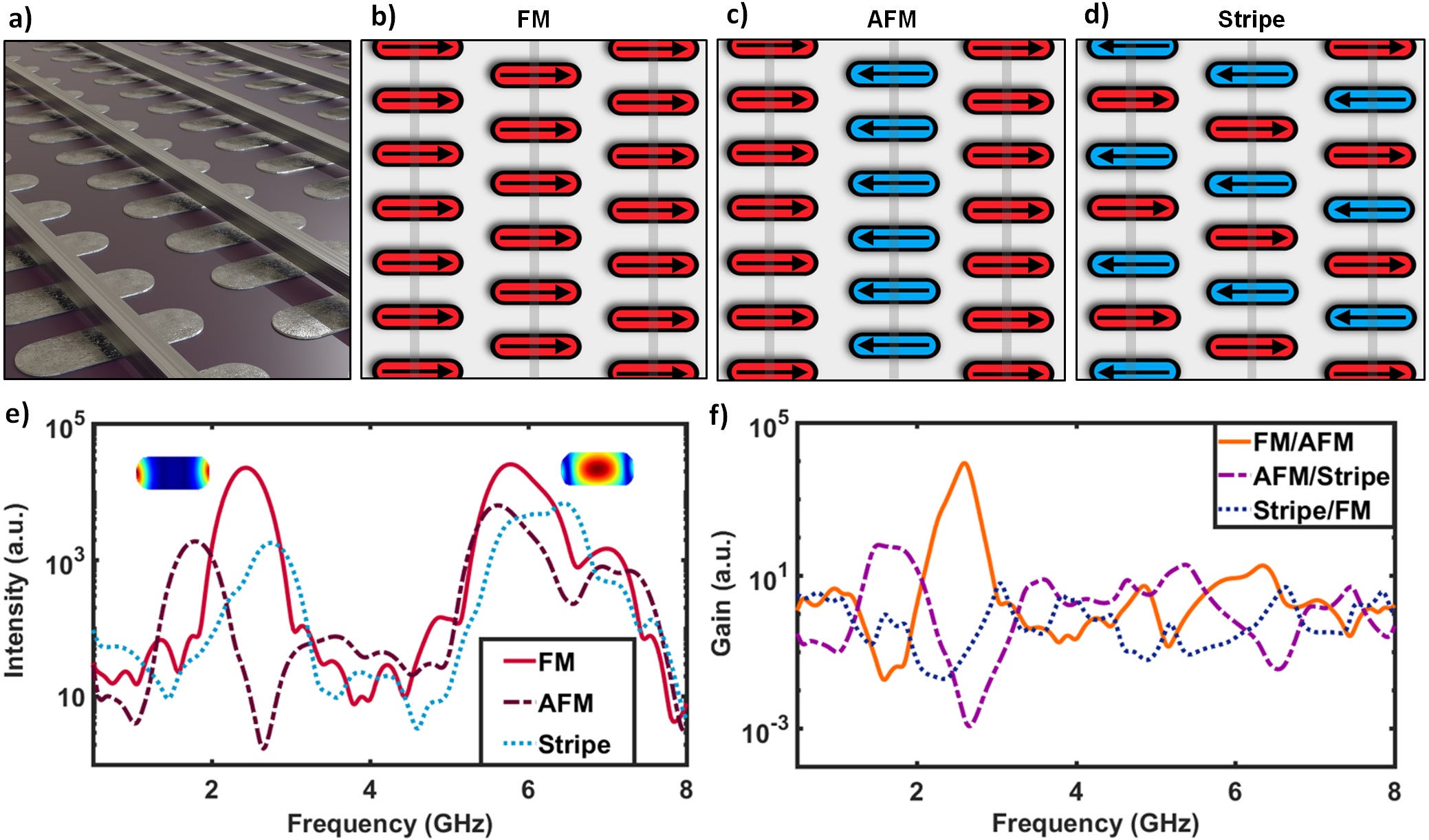}
\caption{a) Schematic of the 2D magnonic crystal incorporating control-nanowires above each column of the nanomagnet array. \\
b)-d) Schematics of the b) FM, c) AFM and d) `stripe' microstates illustrating magnetisation directions of each nanoisland and control-nanowire positions.
e) Spin-wave spectra of ferromagnetic (FM), antiferromagnetic (AFM) and `stripe' microstates on a 4 X 6 nanoisland magnonic crystal. Each state shows two main resonances, an edge-localised mode (1-3 GHz) and a centrally-localised mode (5-8 GHz). Spatial magnon power maps for each mode are displayed next to peaks. \\
f) Spin-wave amplitude gain achieved by transitioning between microstates, i.e. moving from FM to AFM state (solid orange line).}
\label{Fig5} \vspace{-1em}
\end{figure*}

\section*{Reconfigurable magnonic crystals incorporating reversal method}

\subsection*{Reconfigurably-gated magnonic waveguide}

Magnonic device designs hinge on deterministic control of the magnon signal output, either via amplitude or phase modulation. 
Amplitude control has been demonstrated in two-state `transistor'-like devices and via a one-dimensional gateable RMC\cite{haldar2016reconfigurable,chumak2014magnon,cornelissen2018spin,wu2018magnon,cramer2018magnon,fischer2017experimental,chumak2017magnonic,wang2017voltage}. However, gate positions are static - hard-coded at the nanofabrication stage by distinct patterning of `gate' nanoislands relative to the rest of array, and microstate control is restricted to just two of the $2^N$ available states (in an $N$-island array) and reliant on global field sequences. 
These designs function well and are important proofs of concept, but the reliance on global field limits their utility in systems containing magnetically-sensitive states and presents device integration challenges, and static gate-positions limit flexibility and scope. Additionally the number and location of distinctly-patterned gate nanoislands affects the frequency and Q-factor of resonant modes even in the ungated state, causing functional behaviour to vary between arrays with different gate positions or numbers.

Here we present an RMC allowing reconfigurable current-controlled gating with no global field requirement or differential nanoisland patterning. The design comprises a one-dimensional nanoisland array situated above an underlying control-nanowire. Selective nanoisland reversal is achieved by using current-pulses to shift the c-DW position relative to the array as described above. Our reversal scheme allows for a range of gate-types, from single-island gating to more complex multi-island gates offering a range of functional benefits.

The RMC is depicted schematically in fig. \ref{Fig4} a), comprising 25 Py nanoislands of dimensions 350 $\times$ 120 $\times$ 5 nm$^3$ with 60 nm inter-island spacing and 40 $\times$ 40 nm$^2$ cross-section Py control-nanowire along the array long-axis. The first 4 islands are covered by a 550 nm wide Au coplanar waveguide, using pulsed $H_{\mathrm{Oe}}$ to locally excite magnons via which propagate along the RMC via dipolar inter-island interactions. The nanoislands have stable Ising-like magnetisation in the $\pm y$-direction, initialised here with $M = \hat{y}$ referred to as the `ungated' state. 
Fig. \ref{Fig4} h) shows the spatial power map of the magnon edge-mode centered on 2.1 GHz in the ungated array, with good magnon transmission throughout the array. Considering first the fully-selective reversal mode (control nanowire at $h = 20$~nm above nanoislands), figs. \ref{Fig4} b,e,i) show the `single gate' case, where one nanoisland in the array is reversed to $M = -\hat{y}$ via a high-$J$ `write' c-DW traversal. Fig. \ref{Fig4} i) shows magnon transmission strongly attenuated at the reversed gate-island with `on/off' ratios up to 10 observed (fig.\ref{Fig4} e), calculated from the ratio of integrated power at island 20 in the ungated to gated case. 

One benefit of our reversal technique is the freedom to explore gating involving multiple reversed nanoislands. Figs. \ref{Fig4} c,f,j) show one example of this, involving two reversed islands separated by an unreversed island. Termed an `antiferromagnetic'-type gate, on/off ratios are substantially improved to $\sim$30 across a range of gate positions with a peak ratio of 35 at position 5.
While the previous gate types may be accessed in the fully-selective $h = 20$~nm reversal mode, the low-power reversal mode switches all nanoislands traversed by the c-DW so microstates containing isolated reversed islands in otherwise unreversed arrays are not possible. With this in mind, figs. \ref{Fig4} d,g,k) show a $h=10$~nm RMC in a gated state comprising continuously reversed nanoislands one on array side with the other side left unreversed, termed a `domain'-type gate. The presence of the c-DW in the $h=10$ mode introduces a 180$^{\circ}$ DW in the adjacent nanoisland from the influence of $H_{\mathrm{DW}}$, seen in the fig. \ref{Fig4}k) microstate. This is the stable relaxed state, and serves to improve gating efficacy relative to domain-type gates prepared in the $h=20$~nm mode which do not introduce 180 $^{\circ}$ DWs to any nanoisland. The ratios acheived here compare well with existing designs, matching or outperforming the prior art\cite{haldar2016reconfigurable,cramer2018magnon,cornelissen2018spin,wu2018magnon} for a single-island gated one-dimensional array RMC using differential patterning of the gate island. 

\subsection*{Reconfigurable magnonic filter}

The one-dimensional RMC above controls travelling $k \neq 0$ magnon modes. Here we present a second RMC design leveraging the reversal method to control standing-wave $k = 0$ magnons in a two-dimensional nanoisland array.
Manipulating frequency and intensity of standing-wave magnons allows for key functionality including opening and closing specific frequency channels and selective band-pass filtering, with enhanced q-factor and transmission/rejection ratios relative to travelling-wave magnons\cite{semenova2013spin,wang2015tunable,ma2011micromagnetic,kim2009gigahertz,tacchi2015universal,sadovnikov2018spin,mamica2019reversible,mamica2019nonuniform,graczyk2018magnonic}. 

The RMC is depicted schematically in fig. \ref{Fig5} a), comprising adjacent columns of Py nanoislands each with a corresponding DW-carrying control-nanowire. Columns are separated by an gap $g_x$ and islands within a column by gap $g_y$. Adjacent columns are offset in the y-axis by a gap $\frac{g_y}{2}$ to optimise dipolar coupling and hence difference between spectra of distinct microstates. The array considered here has nanoisland dimensions 300 $\times$ 94 $\times$ 5 nm$^3$ and inter-island spacings $g_x = 84$~nm, $g_y = 190$~nm. Dimensions and spacings were optimised for RMC gain functionality. 

Fig. \ref{Fig5} b-d) show three microstates termed `ferromagnetic' (FM) (b), `antiferromagnetic' (AFM) (c) and `stripe' (d), accessible via shuttling c-DWs through control nanowires as described above. 
The RMC was prepared in each microstate and the spin-wave response studied after broadband excitation via an out-of-plane $H_{Ext}$ sinc pulse. Fig. \ref{Fig5}e) shows the resultant spin-wave spectra, with two distinct modes present for each microstate; a nanoisland-edge localised 1-3 GHz mode ($f_{\mathrm{edge}}$ of 2.43 GHz and 1.78 GHz and 2.74 GHz for FM, AFM and stripe respectively),  and a nanoisland-centre localised 5-8 GHz mode, with spatial mode profiles displayed next to each peak. The cause of distinct modes in each microstate is the differing local field $H_{\mathrm{loc}}$ profile from nanoisland stray dipolar fields\cite{gliga2013spectral,zhou2016large,dion2019tunable,arroo2019sculpting,iacocca2016reconfigurable,jungfleisch2017high,jungfleisch2016dynamic,bhat2016magnetization}. These are at their strongest at the nanoisland edges, and it follows that the edge-localised modes display greater microstate sensitivity. This sensitivity allows for considerable mode control, with a frequency shift of $\Delta f = 0.96$~GHz achieved by transitioning between the AFM and stripe states, with the FM state providing an interstitial frequency midpoint.

In addition to frequency shifting, microstate control affords substantial magnon amplitude modulation. Figure \ref{Fig5} f) shows the gain factor obtained by switching between the three microstates, with a gain of $\sim9\times 10^3$ achieved at the FM peak frequency $f =$~2.43 GHz by moving between the FM and AFM states, and a gain of $\sim1\times 10^3$ achieved at the AFM peak frequency $f =$~2.74 GHz by moving between the AFM and stripe states.
The $h = 20$~nm fully-selective mode is capable of accessing the entirety of the RMC's microstate space, the three states presented here are a representative selection of the potential functionality available. States such as `AFM' where entire columns of nanoislands are reversed are accessible in the $h = 10$~nm low-power mode, with the added benefit that control nanowires need only be patterned for the reversed columns - a reduction of 50\% in the AFM case.

The states in fig. \ref{Fig5} are chosen as they have identical $H_{\mathrm{loc}}$ values at each nanoisland (relative to the nanoisland magnetisation) so magnon modes are clearly defined. However, the writing technique enables access to all microstates, including states with different $H_{\mathrm{loc}}$ values at different array positions and hence multiple magnon modes. These mixed-$H_{\mathrm{loc}}$ microstates may be finely tailored, providing control over mode amplitude, frequency, bandwidth and the number of active modes. Detailed discussion of mixed states and their spectra is presented in the supplementary information (figure \ref{2d_RMC_supp}) including the effects of gradual transitions between the FM, AFM and stripe states. 

\section*{Conclusions} 

In this work, we outline a powerful fully solid-state magnetic writing technique, allowing total microstate control without global fields and excellent integration propspects for on-chip systems. The technique allows a class of nanomagnetic systems deriving versatile functionality from a broad range of microstates to become viable candidates for technological advancement, as evidenced by the two novel RMC designs presented here. 

By opening the entire range of microstates for exploration and exploitation, the writing technique described here invites a host of novel nanomagnetic system designs, in addition to the refinement and enhancement of existing designs which are currently limited to narrow regions of microstate space. Functional benefits are offered across diverse applications including neuromorphic logic and superconducting vortex control.

While Py is employed as the ferromagnetic material throughout this work, it is chosen for its ubiquity and familiarity in current nanomagnetic systems and near-zero shape anisotropy rather than ideal magnonic performance. Low-damping ferromagnets such as YIG\cite{chang2014nanometer}, various Heusler alloys\cite{sebastian2012low,nayak2017magnetic} and more simple bimetallic alloys such as CoFe\cite{schoen2016ultra} have superior characteristics for low-loss, longer distance magnon transmission, enhancing functionality of the proposed and related devices incorporating our reversal method. 
We have focused here on magnetic reversal of Ising-like nanoelements. Prior studies have demonstrated injection of 360$^{\circ}$ DWs\cite{gartside2016novel} and Skyrmions using related locally-divergent stray fields (MFM tips in these cases)

\subsection*{Author contributions}
JCG, DMA and WRB conceived the work.\\
DMA wrote initial code for simulation of the reversal method and generation of magnon spectra. \\
SGJ and SYY performed simulations of the reversal method, including the basic functional concept, the two distinct modes and simulations including current-driving DWs using spin-transfer torque. SGJ and SYY further developed and expanded the simulation code with contributions from DMA.\\
JCG, AV, TD and KDS performed simulations of the RMC systems, with JCG, AV and KS working primarily on the 2D RMC and TD working primarily on the 1D RMC. \\
CGI visualisations were designed and rendered by KDS. \\
JCG drafted the manuscript, with contributions from SGJ and SYY in the `reversal method' subsection of the `simulation details' section.\\

\subsection*{Acknowledgements}
This work was supported by the Leverhulme Trust (RPG-2017-257) to WRB. \\
TD and AV were supported by the EPSRC Centre for Doctoral Training in Advanced Characterisation of Materials (Grant No. EP/L015277/1).\\
Simulations were performed on the Imperial College London Research Computing Service\cite{hpc}.
The authors would like to thank Professor Lesley F. Cohen of Imperial College London for enlightening discussion and comments.

\section*{Supplementary Information}

\subsection*{Description of simulation videos}

All nanoislands in the following simulations have dimensions 400 $\times$ 75 $\times$ 5 nm$^3$ and nanowires are 40 $\times$ 40 nm$^2$ cross-section.

\textbf{Supplementary video 1 - Reversal of single nanoisland}

Single nanoisland undergoes magnetic reversal via traversal by current-driven c-DW in suspended nanowire. Nanowire is $h = 10$~nm above nanoisland.\\

\textbf{Supplementary video 2 - Selective reversal of single nanoisland in two-island system}

Right-hand nanoisland in a two nanoisland system is magnetically reversed as in SV1 while left-hand nanoisland is not reversed as the c-DW does not traverse it, demonstrating the local selectivity of the reversal method. \\

\textbf{Supplementary video 3 - Non-selective Oersted field reversal of both nanoislands in a two-island system}

Both nanoislands in a two nanoisland system are simultaneously reversed by the Oersted field caused by current passing through the nanowire. This demonstrates the distinction between the c-DW reversal method described here and conventional, non-selective Oersted field stripline techniques.\\

\textbf{Supplementary video 4 - Fully-selective reversal combining `skip' and `write' events to prepare a `single gate' state}

Reversal method employing varying current-density pulses to achieve combination of `skip' and `write' c-DW traversal events in a four-island system. Nanowire is in $h = 20$~nm fully-selective mode.
Starting from a uniformly-magnetised initial state, a combination of a single `skip' followed by three `write' pulses are used to prepare an arbitrary magnetisation state. The final state corresponds to a `single gate' case, gated at the right-hand island.\\

\textbf{Supplementary video 5 - Fully-selective reversal returning arbitrarily magnetised four-island system to uniform state, then fully-reversed state}

Combination of two `skip' followed by two `write' events returns abitrarily magnetised system to a uniform state. The drive-current polarity is then reversed such that the c-DW motion is right-to-left, and a further combination of two `skip' and two `write' events are employed to prepare a state which is a total magnetic reversal of the initial arbitrary magnetisation state. Nanowire is in $h = 20$~nm fully-selective mode.\\

\textbf{Supplementary video 6 - c-DW skipping across four-island system}

Four `skip' events are combined to move the c-DW across a four island system without reversing any island magnetisation, demonstrating the freedom to reposition the c-DW throughout an array.\\

\subsection*{Gating of magnon centre-mode in low-power domain gate case}

\begin{figure}[htbp]   
\includegraphics[width=1\textwidth]{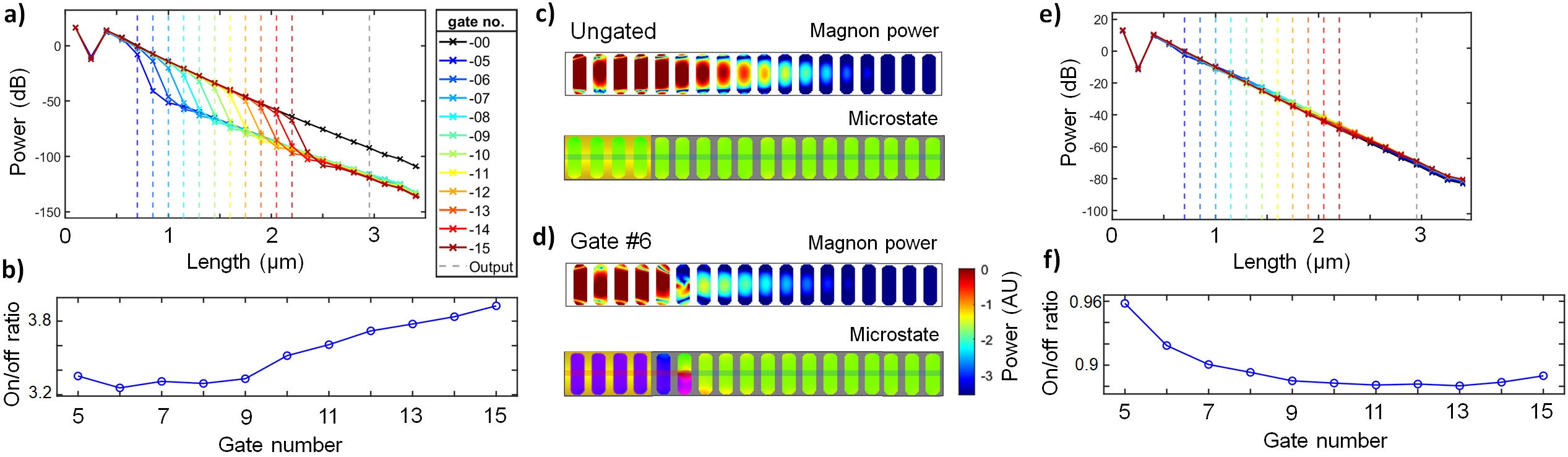}
\caption{a) Bulk-mode spin-wave power vs. array position for `domain' gate type with nanowire in low-power $h = 10$~nm mode. Coplanar microwave waveguide covers islands 1-4. Nanoisland positions denoted by dashed vertical lines with power vs. position traces colour-coded by gate position. \\ 
b) Corresponding magnon `on/off' ratio vs. gate position, calculated from the ratio of integrated power at island 20 in the ungated to gated case. \\ 
c,d) Spatial power map and corresponding microstate of resonant magnon centre-mode for ungated (c) and gate position 6 (d) arrays in low-power mode. Power is normalised to ungated case. Array microstate is shown with green (purple) representing unswitched (switched) islands magnetised in the positive-y (negative-y) direction. Control nanowire is shown over array, coplanar Au waveguide covers islands 1-4.
e,f) Bulk-mode spin-wave power vs. array position (a) and corresponding magnon `on/off' ratio (b) for domain gate case simulated without suspended nanowire and c-DW. In the absence of the c-DW stray field there is no magnetisation distortion in the gate island, and hence very low magnon attenuation is observed.\\}
\label{1d_RMC_supp} 
\end{figure} 

So far the magnons considered have been edge-localised resonant modes. This is because nanoisland edges experience larger variation in local field (and hence larger variation in magnon response) than the centre of islands, as the local stray fields emanate from nanoisland ends. 
In the domain-gate case for the low-power, $h = 10$~nm mode the presence of the c-DW above the gate nanoisland induces significant vortex-like distortion of the island magnetisation, seen in fig. \ref{1d_RMC_supp} d). This distortion is substantial at the nanoisland centre, and as such the centre-localised resonant magnon mode is more strongly attenuated in this gate case than other gate types discussed in this work. Fig. \ref{1d_RMC_supp} e) and f) display the behaviour of the domain-gate simulated with no suspended nanowire and c-DW. In this case there is no magnetisation distortion in the gate-island and hence very low magnon attenuation is observed, demonstrating the utility of the c-DW method to locally distort magnetisation textures as well as offer Ising-like reversal.

\subsection*{Additional microstates of the reconfigurable magnonic filter}

\begin{figure}[htbp]   
\includegraphics[width=1\textwidth]{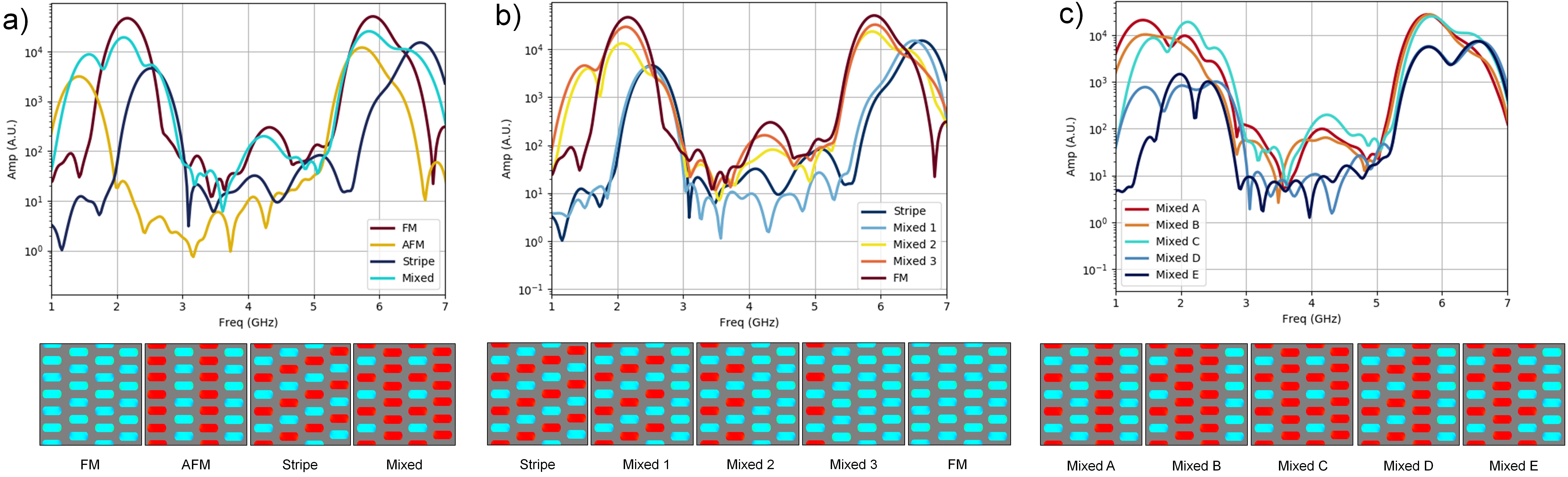}
\caption{Spin-wave spectra of the two-dimensional RMC across a range of microstates. Microstates are shown underneath each spectra with red and blue islands denoting $M = +\hat{x}$ and $M = -\hat{x}$ respectively. A four-column, six-row unit cell is simulated with periodic boundary conditions in the x-direction.\\
a) Spectra of the FM, AFM, stripe and a `mixed' microstate. Mixed microstate comprises elements of the three other states resulting in multiple edge and centre spin-wave modes, especially apparent in the $\sim1-3$~GHz edge-mode region. These modes overlap, supporting a broader range of spin-wave frequencies in the RMC relative to the `pure' microstates and an effective variable-bandwidth filtering mode.\\
b) Spectra showing gradual transition between stripe and FM microstates via three interstitial `mixed' states, switching one column from stripe to FM between each mixed state.\\
Amplitude of the stripe mode is reduced and FM mode is increased as the transition progresses. AFM mode appears during the transition and is most prominent in `mixed 2' state at the transition midpoint. 
Much smoother transitions with finer mode-amplitude control are possible with larger unit-cell arrays.\\ 
c) Spectra of five mixed states, showing the degree to which the relative amplitudes of the FM, AFM and stripe modes may be tuned via microstate design. `Mixed C' state corresponds to the `mixed' state from panel a).  Microstates are named and ordered by the relative intensity of the $\sim1.5$~GHz AFM mode. Again, much finer tuning of the relative mode intensities is possible with larger unit-cell arrays.}
\label{2d_RMC_supp} 
\end{figure}

The FM, AFM and stripe microstates considered so far have symmetrical island magnetisation arrangements, such that the same local dipolar stray field $H_{\mathrm{loc}}$ is felt at each island (relative to its magnetisation). This leads to a single well-defined magnon edge-mode ($\sim1.5-3$~GHz) and centre-mode ($\sim5-7$~GHz) for each microstate. However, the writing technique allows preparation of arbitrary microstates containing a range of $H_{\mathrm{loc}}$ profiles. Figure \ref{2d_RMC_supp} shows a range of these `mixed' microstates and their corresponding spectra. Figure \ref{2d_RMC_supp} a) shows a mixed microstate constituting elements of the previously discussed FM/AFM/stripe states alongside spectra of those states. The mixed state exhibits overlapping modes corresponding to each of the microstates comprising it, resulting in a broader range of permitted magnon frequencies and an effective  variable-bandwidth filtering mode. 

Figure \ref{2d_RMC_supp} b) shows the spectral effects of smoothly transitioning between microstates. The array is prepared in the stripe state and transitioned to the FM state via three interstitial mixed states, with one column of islands reversed between each microstate. The stripe mode smoothly decays away through the first three mixed states, becoming partially obscured from its overlap with the growing FM state from `mixed 2' onwards before abruptly dropping once in the pure FM state. Similarly the FM state can be seen to grow from the mixed 2 state onwards once there are at least two adjacent columns of parallel-magnetised islands. Mixed states 2 and 3 contain aspects of the AFM microstate an as such a related mode at 1.5 GHz is observed. Much smoother microstate transitions are possible with larger unit-cell arrays than the four-column, six-row example used here.

Figure \ref{2d_RMC_supp} c) shows five mixed microstates, each designed comprising aspects of the AFM, FM and stripe states with different relative aspects of each of the three microstates. Mixed states are ordered A)-E) by the relative amount of AFM microstate they contain, which is seen reflected in the AFM magnon mode intensities. Fine control of mode-amplitude tuning is achieved, with enhanced control of the relative mode intensities possible in larger arrays.

\subsection*{Simulation details}

\subsubsection*{Reversal method}

All simulations were performed using MuMax$^3$\cite{vansteenkiste2011mumax, vansteenkiste2014design, leliaert2018fast}.
Magnetic parameters for permalloy (Ni$_{80}$Fe$_{20}$) of $A = 13$~pJ/m, $\alpha =$~0.02 and $M_s = 800\times10^3$~kA/m.  A cell size of $4\times4\times4$\,nm$^{3}$ is used and two numerical solvers are employed at different stages of the simulation - (i) an equilibrium head-to-head TDW is formed in the nanowire by minimising the total energy of the system using the conjugate gradient method and (ii) the temporal evolution of the system is computed by solving the modified version of Landau-Lifshitz-Gilbert (LLG) equation that includes terms for spin-transfer-torque using the Dormand-Prince method (RK45) \cite{vansteenkiste2014design}. The advantage of the Dormand-Prince method is adaptive time steps, dependent on the error magnitude at each iteration, reducing computational intensity.

\begin{figure}[htbp]   
\includegraphics[width=0.45\textwidth]{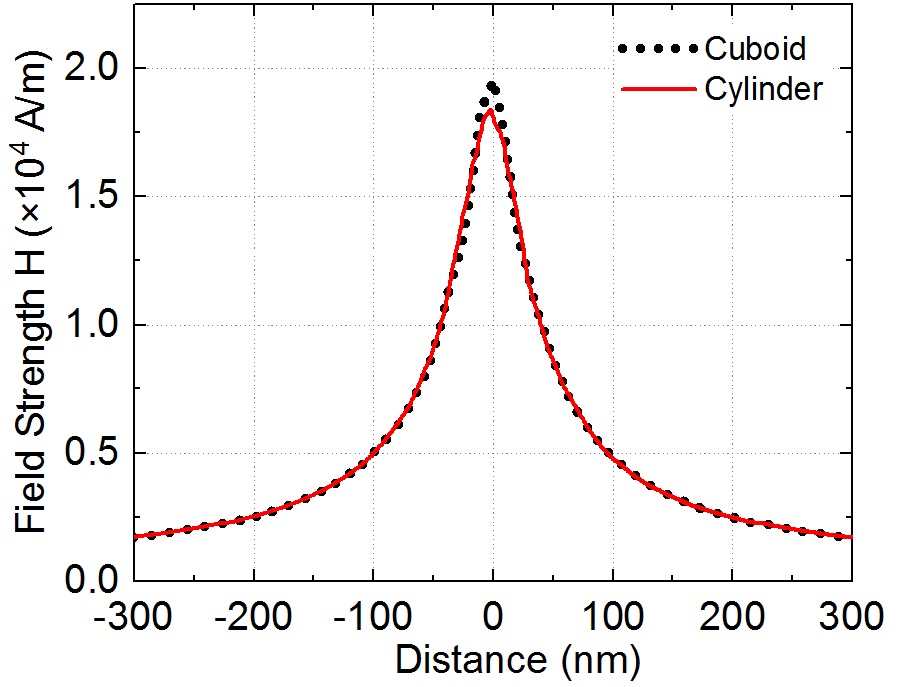}
\caption{Field profiles experienced by a nanoisland that is positioned $h = 10$~nm below an infinitely long cylinder and cuboid. There is a small discrepancy of $0.15\times 10^{4}$Am$^{-1}$ around $x=0$, which can be minimised by adjusting the value of the current density.}
\label{fieldstrength} 
\end{figure}

Boundary conditions are applied to the current-carrying nanowire ends to remove surface charges and associated stray fields, simulating a quasi-infinite wire. A 5 nm window of spins at wire ends are fixed to eliminate turbulent fluctuations when an electric current is applied. 
$H_{\mathrm{Oe}}$ in and around the current-carrying nanowire is simulated by solving Maxwell's equation  $\nabla \times \textbf{B} = \mu_{0}\textbf{J}$, approximating the wire as an infinitely long cylinder and adding the corresponding vector field component to each of the finite-difference cells. 
The validity of the cylindrical-wire assumption was examined using Finite Element Method Magnetics (FEMM)\cite{femm}, with the comparison between $H_{\mathrm{Oe}}$ profiles of cylindrical and cuboid nanowires shown in figure \ref{fieldstrength}. Near identical field profiles are seen, with the small discrepancy for low distances from the nanowire accounted for in the reversal simulations by varying $J$ to match the cuboid case.

\subsubsection*{Reconfigurably-gated magnonic waveguide}
In order to remove stochasticity from nanoisland edge-curl states (i.e. $s-$ and $c$-states) and allow direct comparison between gating types and positions, arrays are relaxed in a 1 mT global field, applied 5$^{\circ}$ above in the positive $x$-direction.

Magnons are excited with a 20 mT sinc-pulse in the $z$-direction applied locally under the Au waveguide, exciting modes up to 20 GHz. The function has a sinusoidal wavelength of 550 nm along the array long-axis, covering the first four nanoislands under the waveguide (see fig. \ref{Fig4} h) for visual of waveguide position). The pulse begins at 50 ps and the magnetisation is recorded every 25 ps for 800 time steps, a total simulation time of 40 ns.

For both RMC simulations, the damping parameter $\alpha$ was reduced to 0.006 for greater correspondence with experiment. Simulation cell sizes of 5 nm were used in all dimensions. 

\subsubsection*{Reconfigurable magnonic filter}
A quasi-infinite array is simulated using periodic boundary conditions, with a two-column, seven island per-column unit cell. Spin-waves are excited by an out-of-plane $H_{Ext}$ sinc pulse applied uniformly across the array, exciting modes between 0.1-25 GHz.
Simulation cell sizes of 3 nm $\times$ 2.2 nm $\times$ 5 nm were used in the $x,y$~and $z$ directions respectively.

Magnons are excited with a 20 mT sinc-pulse in the $z$-direction, exciting modes up to 25 GHz. The function is applied across the entire system. The pulse begins at 50 ps and the magnetisation is recorded every 20 ps for 256 time steps, a total simulation time of 5.12 ns.

\label{Bibliography}
\bibliography{Bibliography.bib}

\end{document}